\documentclass[aip,graphicx]{revtex4-1}
\usepackage[cp1251]{inputenc}
\usepackage[T2A]{fontenc}
\usepackage[english]{babel}
\usepackage{amssymb,latexsym,amsmath,amscd}
\usepackage{graphicx,color,framed}
\begin{document}
\selectlanguage{english}

\title{Polarizable polymer chain under external electric field in a dilute polymer solution}

\author{\firstname{Yu.~A.} \surname{Budkov}}
\email[]{urabudkov@rambler.ru}
\affiliation{G.A. Krestov Institute of Solution Chemistry of the Russian Academy of Sciences, Laboratory of NMR spectroscopy and numerical investigations of liquids, Ivanovo, Russia}
\affiliation{National Research University Higher School of Economics, Department of Applied Mathematics, Moscow, Russia}

\author{\firstname{ A.~L.} \surname{Kolesnikov}}
\affiliation{Institut f\"{u}r Nichtklassische Chemie e.V., Universitat Leipzig, Leipzig, Germany}

\author{\firstname{ M.~G.} \surname{Kiselev}}
\affiliation{G.A. Krestov Institute of Solution Chemistry of the Russian Academy of Sciences, Laboratory of NMR spectroscopy and numerical investigations of liquids, Ivanovo, Russia}
\begin{abstract}
We study the conformational behavior of polarizable polymer chain under an external homogeneous electric field within the Flory type self-consistent field theory. We consider the influence of electric field on the polymer coil as well as on the polymer globule. We show that when the polymer chain conformation is a coil, application of external electric field leads to its additional swelling. However, when the polymer conformation is a globule, a sufficiently strong field can induce a globule-coil transition. We show that such $"$field-induced$"$ globule-coil transition at the sufficiently small monomer polarizabilities goes quite smoothly. On the contrary, when the monomer polarizability exceeds a certain threshold value, the globule-coil transition occurs as a dramatic expansion in the regime of first-order phase transition. The developed theoretical model can be applied to predicting polymer globule density change under external electric field in order to provide more efficient processes of polymer functionalization, such as sorption, dyeing, chemical modification, etc.
\end{abstract}

\maketitle
\section{Introduction}

A coil-globule transition in dilute polymer solutions is one of the most famous phenomena in polymer physics. The importance of this phenomenon is explained by the fact that coil-globule transition plays a crucial role in recent technological advances from targeted delivery of drug molecules to their encapsulation in a polymer coil \cite{DrugDeliveryReview1,DrugDeliveryReview2,DrugDeliveryReview3}. On the other hand, understanding of the conformational transitions of the single polymer chain in the solution is basis of understanding of the thermodynamic and structural properties of the so-called soft active materials, including dielectric elastomers \cite{Diel_elast}, elastomeric gels \cite{Doi}, polyelectrolytes \cite{Dobrynin1,Dobrynin2,Holm_Review,Erukhimovich1,Baeurle2009,Budkov_2011,Nogovitsin2012,Budkov_2013,Budkov2014,Budkov2015,Kolesnikov2014,Budkov2015_2}, copolymers \cite{Erukhimovich2,Wang}, pH-sensitive hydrogels \cite{Marcombe}, and temperature-sensitive hydrogels \cite{Cai}. Today, the great efforts taken to develop a coil-globule transition theory have made a great contribution to an understanding of this phenomenon. The coil-globule transition has been considered within a simple theory of self-consistent field as well as a sophisticated field-theoretic formalism \cite{Grosberg,Birshtein_1,Moore,deGennes_collaps,Sanchez,Khohlov,Brilliantov_collaps,Halperin,Budkov1,Budkov2,Budkov3,Odagiri2015}. Most of these studies are based on the simple idea that the polymer coil shrinks lead to a collapse, when the solvent becomes poorer. It should be noted that the coil-globule transition usually occurs as first- or second-order phase transitions \cite{Khohlov_book}.

The possibilities to control the conformational transition by an external stimulus have attracted great attention in chemistry, biology, and material science. One of the most interesting external stimuli for different industrial applications enabling the control of the conformations of polymers, such as polyelectrolytes, dielectric elastomers, liquid-crystalline polymers, etc., is application of an external electric field. A lot of theoretical studies of the conformational behavior of single polyelectrolyte chain in the external homogeneous electric field have been published \cite{Muthukumar2,Podgornik,Borisov,Joanny,Netz,Brilliantov,Seidel}. In case of single polyelectrolyte chain, its response to the electric field is achieved by the presence of an electric charge on the polymer backbone. However, there is a wide class of neutral dielectric polymers with permanent dipoles or a molecular polarizability of monomers that also make it possible to act on the polymer chain conformation by electric field application. Up to now, several theoretical works have been published where  thermodynamic and structural properties of dielectric polymers in the bulk solution have been discussed \cite{Muthu_1996,Podgornik_2004,Kumar_2009,Dean_2012,Kumar_2014}. Nevertheless, to the best of our knowledge, until now there have been no the systematic theoretical studies of conformational behavior of a flexible dielectric polymer chain under the external electric field even at the level of mean-field approximation.

In this short communication we undertake the first effort in this direction. We address the theoretical study of the polarizable polymer chain dissolved in a dielectric solvent under external homogeneous electric field within the Flory type self-consistent field theory \cite{Flory_book}. We study the influence of the external electric field on the polymer coil as well as on the polymer globule. We show that applying electric field leads to an additional swelling of the polymer coil. However, when the polymer chain conformation is a globule, we find out that a sufficiently large electric field can provoke globule-coil transition.

\section{Theory}
Let us consider a polarizable polymer chain dissolved in a solvent which we shall model as a continuous dielectric media with the dielectric permittivity $\varepsilon_{s}$.  We assume that the polymer chain has a degree of polymerization $N$ and that each monomer has a molecular polarizability $\gamma$. We also assume that the polymer chain is placed in a homogeneous electric field $\bold{E}$. To study the conformations of the polymer chain in the external electric field, we shall use the simple Flory type \cite{Flory_book} self-consistent field theory, considering the radius of gyration $R_{g}$ as a single order parameter. We also assume that the polymer chain occupies the volume which can be expressed through the gyration volume $V_{g}=4\pi R_{g}^3/3$.  Thus, a total free energy of the polymer chain can be written in the following form:
\begin{equation}
\mathcal{F}(R_{g})=\mathcal{F}_{conf}(R_{g})+\mathcal{F}_{vol}(R_{g})+\mathcal{F}_{el}(R_{g}),
\end{equation}
where $\mathcal{F}_{conf}(R_{g})$ is the free energy of the ideal polymer chain which can be calculated by the following interpolation formula \cite{Fixman,Grosberg,Khohlov_book}
\begin{equation}
\label{eq:conf}
\mathcal{F}_{conf}(R_{g})=\frac{9}{4}k_{B}T\left(\alpha^{2}+\frac{1}{\alpha^2}\right),
\end{equation}
where $\alpha=R_{g}/R_{0g}$  is the expansion factor, $R_{0g}^2=Nb^2/6$ is the mean-square radius of gyration of the ideal Gaussian polymer chain, $b$ is the Kuhn length of the segment, $k_{B}$ is the Boltzmann constant, $T$ is the absolute temperature;
the contribution of monomer volume interactions to the total free energy takes the following form
\begin{equation}
\label{eq:vol}
\mathcal{F}_{vol}(R_{g})=k_{B}T\left(\frac{N^2B}{2V_{g}}+\frac{N^3C}{6V_{g}^2}\right),
\end{equation}
where $B$ and $C$ are the second and third virial coefficients of the  monomer-monomer interactions, respectively;
the electrostatic contribution can be written as 
\begin{equation}
\label{eq:electr}
\mathcal{F}_{el}(R_{g})=V_{g}\left(\frac{\varepsilon_{p}\mathcal{E}^2}{8\pi}-\frac{\varepsilon_{s}E^2}{8\pi}\right),
\end{equation}
where $\varepsilon_{p}$ is the effective dielectric permittivity inside the polymer volume, which can be determined in the mean-field approximation as \cite{Budkov}
\begin{equation}
\label{eq:permit}
\varepsilon_{p}=\varepsilon_{s}+\frac{4\pi\gamma N}{V_{g}},
\end{equation}
$\mathcal{E}=\varepsilon_{s}E/\varepsilon_{p}$ is the electric field inside the polymer volume, $\varepsilon_{s}$ is the solvent dielectric permittivity, $\gamma$ is the molecular polarizability of monomers, 
and $E=|\bold{E}|$ is the absolute value of the external electric field.

Using expression (\ref{eq:permit}) and the expression for the local electric field $\mathcal{E}=\varepsilon_{s}E/\varepsilon_{p}$ inside the gyration volume, after some calculations we obtain the following expression for the total free energy of the polymer chain in the solution under external electric field:
\begin{equation}
\label{eq:total}
\frac{\mathcal{F}}{k_{B}T}=\frac{9}{4}\left(\alpha^{2}+\frac{1}{\alpha^2}\right)+\frac{N^2B}{2V_{g}}+\frac{N^3C}{6V_{g}^2}-\frac{N\varepsilon_{s}\gamma E^2}{2k_{B}T\left(\varepsilon_{s}+\frac{4\pi\gamma N}{V_{g}}\right)}.
\end{equation}

Minimizing total free energy (\ref{eq:total}) with respect to the expansion factor $\alpha$, after some algebra we arrive at
\begin{equation}
\label{eq:alpha}
\alpha^5-\alpha=\frac{3\sqrt{6}}{2\pi b^3}B\sqrt{N}+\frac{27C}{\pi^2\alpha^3b^6}+\frac{6\sqrt{6}\gamma^2\varepsilon_{s}\sqrt{N}E^2}{k_{B}Tb^3\left(\varepsilon_{s}+\frac{18\sqrt{6}\gamma}{\alpha^3b^3\sqrt{N}}\right)^2}.
\end{equation}
As is seen from eq. (\ref{eq:alpha}), the interaction of the induced dipoles of the polymer chain with the external electric field leads to the macromolecule swelling. The latter is  well known in thermodynamics of dielectrics as an electrostriction effect \cite{Landau_VIII}.

\subsection{Regime of good solvent}
Eq. (\ref{eq:alpha}) allows us to study the influence of the electric field on the conformation of a polarizable polymer chain. Firstly, let us consider the case of good solvent when the polymer chain without the electric field is in coil conformation. 
In this case $B>0$ and the expansion factor $\alpha\gg 1$, so that the second term in the right-hand side of (\ref{eq:alpha}) (that is related to the triple monomer correlation) can be safely omitted. 
Then we get the following simplified equation for the expansion factor
\begin{equation}
\label{eq:alpha2}
\alpha^5-\alpha=\frac{3\sqrt{6}}{2\pi b^3 }\sqrt{N}\left(B+\frac{4\pi\gamma^2E^2}{k_{B}T\varepsilon_{s}}\right).
\end{equation}

As one can see from (\ref{eq:alpha2}), the presence of electric field inside the polymer coil only leads to an effective increase in the second virial coefficient and, consequently, its additional swelling. In case of strong electric field ($E\rightarrow\infty$), we get the following evaluations for the expansion factor and the radius of gyration:
\begin{equation}
\alpha\simeq \left(\frac{6\sqrt{6}\gamma^2}{k_{B}Tb^3\varepsilon_{s}}\right)^{1/5}E^{2/5}N^{1/10},~~~ R_{g}\simeq \left(\frac{b^2\gamma^2}{6k_{B}T\varepsilon_{s}}\right)^{1/5}E^{2/5}N^{3/5}.
\end{equation}

It should be noted that in case of strong electric field the size of the polymer coil is determined by the monomer polarizability, the temperature, the solvent dielectric permittivity, and the external electric field, but it does not depend on the parameters of monomer pair interactions.

\subsection{Regime of poor solvent}

Let us now consider a case of a poor solvent when the polymer chain without the electric field is in the globule conformation. In this regime, the second virial coefficient is a negative value ($B<0$), while the expansion factor $\alpha\ll 1$.  As is well known, in order to describe the conformational behavior of a polymer chain in the regime of poor solvent, we have to take into account the triple monomer correlations, leaving the contribution in (\ref{eq:alpha}) which is related to the third virial coefficient \cite{Khohlov_book}. In the case when $E=0$, the expansion factor can be evaluated as $\alpha\simeq \left(3\sqrt{6}C/\pi b^3|B|\right)^{1/3}N^{-1/6}$. Further, neglecting the terms in the left-hand side of  (\ref{eq:alpha}) and introducing a notation $\alpha=sN^{-1/6}$, we obtain the following equation for parameter $s$:
\begin{equation}
\label{eq:s}
B+\frac{4\pi\gamma^2E^2\varepsilon_{s}}{k_{B}T\left(\varepsilon_{s}+\frac{18\sqrt{6}\gamma}{s^3b^3}\right)^2}+\frac{3\sqrt{6}C}{\pi s^3b^3}=0.
\end{equation}

At a sufficiently small electric field the equation (\ref{eq:s}) can be solved as
\begin{equation}
\label{eq:s2}
s\simeq\left(\frac{3\sqrt{6}C}{\pi b^3|B_{eff}(E)|}\right)^{1/3},
\end{equation}
where an effective second virial coefficient $B_{eff}(E)=B+4\pi\gamma^2\varepsilon_{s}E^2/k_{B}T\left(\varepsilon_{s}+\frac{6\pi\gamma |B|}{C}\right)^2$ is introduced.
As it follows from relation (\ref{eq:s2}), as the electric field grows, the radius of gyration monotonically increases too. Solution (\ref{eq:s2}) allows us to calculate the globule density
\begin{equation}
\label{eq:glob_dens}
\rho_{g}=\frac{N}{V_{g}}\simeq \frac{3|B|}{2C}-\frac{6\pi\gamma^2\varepsilon_{s}E^2}{Ck_{B}T\left(\varepsilon_{s}+\frac{6\pi\gamma |B|}{C}\right)^2}.
\end{equation}
The first term in the right-hand side of eq. (\ref{eq:glob_dens}) determines the globule density at zero electric field. The second term is the first correction on the electric field $E$. 
At a further electric field increase the globule-coil transition may take place, which will be discussed detail in the next subsection.

\subsection{Field-induced globule-coil transition}

To proceed the numerical analysis of the above-mentioned in previous section field-induced globule-coil transition, we introduce the following dimensionless variables: $\tilde{E}=E\sqrt{\varepsilon_{s}b^3/k_{B}T}$, $\tilde{B}=Bb^{-3}$, $\tilde{C}=Cb^{-6}$, and $\tilde{\gamma}=\gamma b^{-3}/\varepsilon_{s}$. Thus, basic equation (\ref{eq:alpha}) can be rewritten as
\begin{equation}
\label{eq:alpha3}
\alpha^5-\alpha=\frac{3\sqrt{6}}{2\pi}\sqrt{N}\left(\tilde{B}+\frac{4\pi \tilde{\gamma}^2\tilde{E}^2}{\left(1+\frac{18\sqrt{6}\tilde{\gamma}}{\alpha^3\sqrt{N}}\right)^2}\right)+\frac{27\tilde{C}}{\pi^2\alpha^3}.
\end{equation}
We choose parameters $\tilde{B}$ and $\tilde{C}$ so that polymer is in the globular conformation at zero electric field, namely $\tilde{B}=-0.25$ and $\tilde{C}=0.4$.

Fig. 1 demonstrates the dependencies of expansion factor $\alpha$ on the external electric field $\tilde{E}$ at different monomer polarizabilities $\tilde{\gamma}$, obtained by solving eq. (\ref{eq:alpha3}).  As is seen, at the sufficiently small monomer polarizabilities, the electric field increase causes the expansion factor to grow smoothly. It should be noted that in the limit of the infinitely long polymer chain ($N\rightarrow\infty$) at the sufficiently small monomer polarizability, the globule-coil transition occurs as a second-order phase transition. However, above some critical value $\tilde{\gamma}_{c}$ of the monomer polarizability, the expansion factor dependence on the electric field looks like an "S-shaped" curve. The latter means that the globule-coil transition at large monomer polarizabilities occurs as a first-order phase transition. The regions of metastable states and of unstable states are marked by the red and blue lines, respectively. It should be noted that in reality at certain electric field an abrupt increase (black vertical lines) in the expansion factor should take place. We would also like to stress that the expansion factor values between which the abrupt increases occur were found to be the points of global minima of the total free energy. Fig.1 also illustrates the binodal with the critical point (green line).

\begin{figure}
\centerline{\includegraphics[scale = 1]{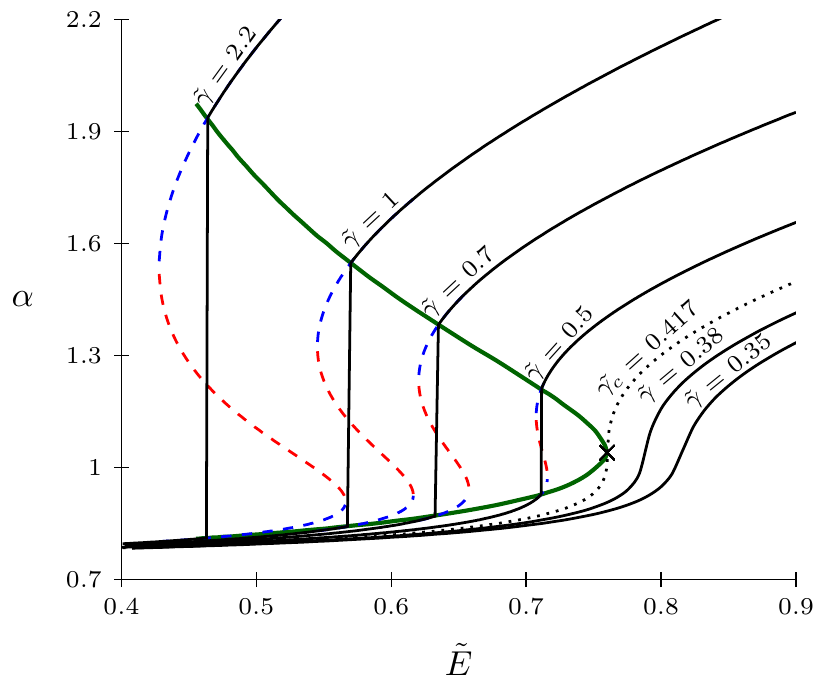}}
\caption{The phase diagram is presented in coordinates $\alpha$ vs $\tilde{E}$ for the polarizable polymer chain. At the sufficiently small monomer polarizabilities the electric field increase causes the expansion factor to grow smoothly. Above some critical value $\tilde{\gamma}_{c}$ of the monomer polarizability the expansion factor dependence on the electric field looks like an "S-shaped" curve that indicates on the first-order phase transition. The regions of metastable states (blue lines) and of unstable states (red lines) are marked. The binodal is marked by the green line. The data are shown for $\tilde{B}=-0.25$, $\tilde{C}=0.4$, $N=100$.}
\label{fig.1}
\end{figure}

\section{Concluding remarks}

In this communication we have presented a simple self-consistent field theory of the conformational behavior of polarizable polymer chain in the dilute polymer solution under external homogeneous electric field. We have shown that in the regime of good solvent, when the polymer chain adopts a coil conformation, application of the electric field leads to additional swelling of the polymer coil. However, in the regime of poor solvent, when the polymer chain conformation is a globule, we have obtained that the electric field can induce a globule-coil transition in two qualitatively different ways. Namely, at the sufficiently small monomer polarizability the globule-coil transition occurs quite smoothly. On the contrary, when the monomer polarizability exceeds a certain value, the globule-coil transition occurs as a first-order phase transition, i.e. as a discontinuous change of the radius of gyration.

We would like to stress that within the present model we assume the isotropic dielectric response of the polymer chain. Such simplification is due to the fact that in this study we would like to discuss only the principal possibility to change polarizable polymer chain conformation by electric field at simplest level of theory, leaving more rigorous consideration with an account of the response anisotropy for future research. Nevertheless, we believe that accounting for the dielectric response anisotropy could give some new interesting regimes.

In conclusion, we would like to speculate on the possible applications of the discussed phenomena. Firstly, we believe that they might offer an additional possibility to control polymer conformation in the solutions for encapsulation of drug molecules into a polymer coil and their subsequent targeted delivery. Secondly, application of electric field, decreasing the polymer density, can help to provide more efficient  processes  of polymer functionalization, such as sorption, dyeing, chemical modification, etc.

\begin{acknowledgments}
This work was supported by grant from the President of the Russian Federation (No MK-2823.2015.3).
\end{acknowledgments}

\newpage


\begin{thebibliography}{99}

\bibitem{DrugDeliveryReview1}
{Kost J., Langer R.} Advanced Drug Delivery Reviews $\bold{46}$, 125 (2001).

\bibitem{DrugDeliveryReview2}
{Priya Bawa, Viness Pillay, Yahya E Choonara and Lisa C du Toit}  Biomed. Mater. $\bold{4}$, 022001 (2009). 

\bibitem{DrugDeliveryReview3}
{ Fitzpatrick S.D., Lindsay E.F., Thakur A., et.al.}  Expert Review of Medical Devices  $\bold{9}$ (4), 339 (2012).

\bibitem{Diel_elast}
{Zhigang Suo} Acta Mechanica Solida Sinica $\bold{23}$ (6), 549 (2010).

\bibitem{Doi}
{Doi M.} Journal of the Physical Society of Japan $\bold{78}$, 052001 (2009).

\bibitem{Dobrynin1}
{A.V. Dobrynin, M. Rubinstein} Prog. Polym. Sci. $\bold{30}$, 1049 (2005).

\bibitem{Dobrynin2}
{A.V. Dobrynin} Curr. Opin. Colloid Int. Science $\bold{13}$, 376 (2008).

\bibitem{Holm_Review}
{Holm C., Joanny J.F., Kremer K., Netz R.R., Reineker P., Seidel C., Vilgis T.A., Winkler R.G.} Adv. Polym. Sci. $\bold{66}$, 67 (2004).

\bibitem{Erukhimovich1}
{V. Yu. Boryu, I. Ya. Erukhimovich} Macromolecules $\bold{21}$ (11), 3240 (1988).

\bibitem{Baeurle2009}
{S.A. Baeurle, M.G. Kiselev, E.S. Makarova, and E.A. Nogovitsin} Polymer $\bold{50}$ (7), 1805 (2009).

\bibitem{Budkov_2011}
{Nogovitsyn E. A., Budkov Yu. A.}  Russ. J. Phys. Chem. A 2011, $\bold{85}$, 1363-1368.

\bibitem{Nogovitsin2012}
{Nogovitsin E.A., Budkov Yu.A.} Physica A $\bold{391}$, 2507 (2012).

\bibitem{Budkov_2013}
{Budkov Yu. A., Nogovitstyn E. A., Kiselev M. G.} Russ. J. Phys. Chem. A 2013, 87, 638644.

\bibitem{Budkov2014}
{Budkov Yu.A., Kolesnikov A.L., Nogovitsyn E.A., Kiselev M.G.} Polymer Science, Ser. A $\bold{56}$ (5), 697 (2014).

\bibitem{Budkov2015}
{Budkov Yu.A., Kolesnikov A.L., Georgi N., Nogovitsyn E.A., Kiselev M.G.} J. Chem. Phys. $\bold{142}$, 174901 (2015).

\bibitem{Budkov2015_2}
{Budkov Yu.A., Kolesnikov A.L., Georgi N., Nogovitsyn E.A., Kiselev M.G.} J. Chem. Phys. $\bold{143}$, 189903 (2015).

\bibitem{Kolesnikov2014}
{Kolesnikov A.L., Budkov Yu.A., Nogovitsyn E.A.} J. Phys. Chem. B $\bold{118}$, 13037 (2014).

\bibitem{Erukhimovich2}
{Thorsten Goldacker, Volker Abetz, Reimund Stadler, Igor Erukhimovich, and Ludwik Leibler} Nature $\bold{398}$, 137 (1999).

\bibitem{Wang}
{Liquan Wang, Jiaping Lin, Xu Zhang} Polymer $\bold{54}$, 3427 (2013).

\bibitem{Marcombe}
{Marcombe R., Cai S.Q., Hong W., Zhao X.H., Lapusta Y. and Suo Z.G.} Soft Matter $\bold{6}$, 784 (2010).

\bibitem{Cai}
{Cai S.Q. and Suo Z.G.} Journal of the Mechanics and Physics of Solids $\bold{59}$ (11), 2259 (2011).

\bibitem{Grosberg}
{Grosberg A.Yu., Kuznetsov D.V.}  Macromolecules $\bold{25}$, 1970 (1992).

\bibitem{Birshtein_1}
{Birshtein T.M., Pryamitsyn V.A.}  Macromolecules $\bold{24}$, 1554 (1991).

\bibitem{Moore}
{Moore M.A.} J. Phys. A: Math. Gen. $\bold{10}$ (2), 305 (1977). 

\bibitem{deGennes_collaps}
{de Gennes P.G.} Le Journal De Physique - Letters. $\bold{36}$ (2),  P. L-55 (1975).

\bibitem{Sanchez} 
{Sanchez I.C.}  Macromolecules $\bold{12}$ (5), 980 (1979).

\bibitem{Khohlov}
{Khokhlov A.R.}  Physica A $\bold{105}$,  P. 357 (1981).

\bibitem{Brilliantov_collaps}
{Brilliantov N.V., Kuznetsov D.V., Klein R.} Phys. Rev. Lett. $\bold{81}$ (7), 1433 (1988).

\bibitem{Halperin}
{A. Halperin and E. B. Zhulina}  Euro. Phys. Lett. $\bold{15}$ (4), 417 (1991).

\bibitem{Budkov1}
{Budkov Yu.A., Kolesnikov A.L., Georgi N., and Kiselev M.G.} J. Chem. Phys. $\bold{141}$, 014902 (2014).

\bibitem{Budkov2}
{Budkov Yu.A., Vyalov I.I., Kolesnikov A.L., et.al.} J. Chem. Phys. $\bold{141}$, 204904 (2014).

\bibitem{Budkov3}
{Budkov Yu.A., Kolesnikov A.L., Georgi N., Kiselev M.G.} Euro. Phys. Lett. $\bold{109}$, 36005 (2015).

\bibitem{Odagiri2015}
{Kenta Odagiri and Kazuhiko Seki} J. Chem. Phys. $\bold{143}$, 134903 (2015).

\bibitem{Khohlov_book}
{A.Yu. Grosberg and A. R. Khokhlov} {\sl Statistical Physics of Macromolecules} (AIP, New York, 1994).

\bibitem{Muthukumar2}
{Muthukumar M.} J. Chem. Phys. $\bold{86}$, 7230 (1987).

\bibitem{Podgornik}
{Podgornik R. and Jonsson B.}  Euro. Phys. Lett. $\bold{24}$, 501 (1993).

\bibitem{Borisov}
{Borisov O.V., Zhulina E.B. and Birshtein T.M.} J. Phys. II $\bold{4}$, 913 (1994).

\bibitem{Joanny}
{Chatellier X. and Joanny J.-F.} Phys. Rev. E $\bold{57}$, 6923 (1998).

\bibitem{Netz}
{Netz R.R.} J. Phys. Chem. B $\bold{107}$, 8208 (2003).

\bibitem{Brilliantov}
{Brilliantov N.V. and Seidel C.} Euro. Phys. Lett. $\bold{97}$, 28006 (2012).

\bibitem{Seidel}
{Seidel C., Budkov Yu.A., Brilliantov N.V. } Nanoengineering and Nanosystems $\bold{227}$, 142 (2013).

\bibitem{Muthu_1996}
{M. Muthukumar} J. Chem. Phys. $\bold{104}$, 691 (1996).

\bibitem{Podgornik_2004}
{Rudi Podgornik} Phys. Rev. E $\bold{70}$, 031801 (2004).

\bibitem{Kumar_2009}
{Rajeev Kumar and Glenn H. Fredrickson} J. Chem. Phys. $\bold{131}$, 104901 (2009).

\bibitem{Dean_2012}
{David S. Dean and Rudolf Podgornik}  J. Chem. Phys. $\bold{136}$, 154905 (2012).

\bibitem{Kumar_2014}
{Rajeev Kumar, Bobby G. Sumpter, and M. Muthukumar} Macromolecules $\bold{47}$, 6491 (2014).

\bibitem{Flory_book}
{Flory P.} {\sl Statistical Mechanics of Chain Molecules} (Wiley-Interscience, New York, 1969).

\bibitem{Fixman}
{Fixman M.} J. Chem. Phys. $\bold{36}$ (2), 306 (1962).

\bibitem{Budkov}
{Yu.A. Budkov, A.L. Kolesnikov, and M.G. Kiselev} Euro. Phys. Lett. $\bold{111}$, 28002 (2015).

\bibitem{Landau_VIII}
{Landau L.D., Lifshitz E.M.}  {\sl Electrodynamics of Continuous Media V. 8, A Course of Theoretical Physics} (Pergamon Press, Oxford, 1960).


\end{thebibliography}
\end{document}